# Multiple X-ray bursts from long discharges in air

C.V. Nguyen[1], A.P.J. van Deursen[1] and U. Ebert[2]

[1] Department of Electrical Engineering, Eindhoven University of Technology, POBox. 513, NL-5600 MB Eindhoven, The Netherlands
[2] Department of Applied Physics, Eindhoven University of Technology and CWI, Centre for Mathematics and Informatics, POBox 94079, NL-1090 GB Amsterdam, The Netherlands

**Abstract**
A lightning surge generator generates a high voltage surge with 1.2 µs rise time. The generator fed a spark gap of two pointed electrodes at 0.7 to 1.2 m distances. Gap breakdown occurred between 0.1 and 3 µs after the maximum generator voltage of approximately 850 kV. Various scintillator detectors with different response time recorded bursts of hard radiation in nearly all surges. The bursts were detected over the time span between approximately half of the maximum surge voltage and full gap breakdown. The consistent timing of the bursts with the high-voltage surge excluded background radiation as source for the high intensity pulses. In spite of the symmetry of the gap, negative surges produced more intense radiation than positive. This has been attributed to additional positive discharges from the measurement cabinet which occurred for negative surges. Some hard radiation signals were equivalent to several MeV. Pile-up occurs of lesser energy X-ray quanta, but still with a large fraction of these with an energy of the order of 100 keV. The bursts occurred within the 4 ns time resolution of the fastest detector. The relation between the energy of the X-ray quanta and the signal from the scintillation detector is quite complicated, as shown by the measurements.

**1. Introduction**
As early as 1924 Wilson stated [1] that 'By its accelerating action on particles the electric field of a thundercloud may produce extremely penetrating corpuscular radiation.' High-energy radiation has indeed been associated with lightning, as it has been observed in measurements from space [2,3], in balloon flights [4] and at surface level [5,6]. A plausible explanation is Bremsstrahlung due to runaway electrons at high altitudes with energies above 100 keV [7-9]. The lightning natural occurrence is quite random in time and position, making it a difficult object for detailed study at high altitudes. Energetic radiation has also been observed with discharges in gases at STP[1]. An overview of recent theoretical and experimental studies of long sparks and their formation is given in [10-12]. However, the processes to accelerate electrons in STP air up to the required high energies are not really understood yet.

We studied sparks of the order of 1 m length in the laboratory, and focused our attention to where and when in the developing discharge channel the high energy radiation is generated. In comparison to earlier work [13] we added a larger number of measurements to allow a statistical analysis. The setup will be presented in detail, because it strongly influenced the timing, intensity and position of the hard X-ray generation, as will be discussed in Section 3. A recent paper [14] presents results on a careful experiment, similar to but independent of ours; we discuss similarities and differences in Section 4.

The scintillator detector used in most of our measurements has a good energy resolution for single γ-ray quanta in the photopeak. However, the relation between detector output and X-ray or γ-quantum energy is generally not straightforward, as will be discussed in Section 2.1. We tried to resolve the ambiguity by placing lead or aluminum absorbers of various thicknesses in front of the detector. Ultimately, a theoretical description of the developing discharge should provide an energy distribution of electrons. The Bremsstrahlung process further complicates the relation between electron and radiation energy.

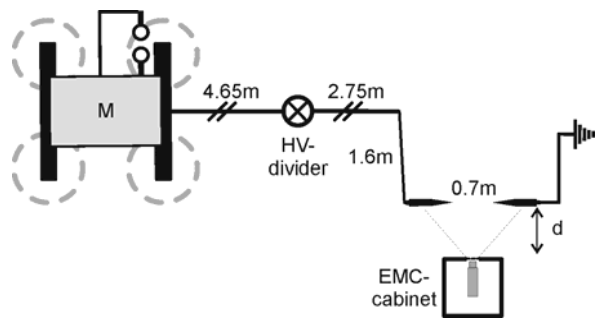

**Figure 1:** Floor plan of the setup, showing position of Marx generator (M), HV-divider, 0.7 m spark gap and EMC cabinet with detector at the distance *d* from the gap.

**2. Experimental setup**
For our experiments we used the 2 MV twelve stage Marx [15] generator in the High-Voltage Laboratory at Eindhoven University of Technology. The voltage waveform of the unloaded generator is a standardized lightning surge with 1.2 µs rise time and 50 µs decay to half-maximum. The surge amplitude and polarity can be chosen. The 9 m tall 1:2000 high-voltage (HV) divider is a part of the wave shaping circuit.

---

[1] Standard temperature and pressure, 21 degrees Celsius and 1 atmosphere.

Figure 1 shows the floor plan of the setup. A spark gap consisting of two pointed aluminum electrodes (cone angle 21 degrees, tip radius about 1 mm) were placed on insulating stands at 2 m above the floor. One electrode was connected to the divider HV end, the other to the conducting floor. The 0.7 m distance between the tips typically used ensured full gap breakdown at approximately 1 MV surge voltage within one or a few microseconds after the maximum voltage $V_{max}$. Of course this delay depended on the electrode distance and Marx generator setting.

A grounded EMC[2] cabinet [16] faced the spark at the distance of $d = 0.8$ m and more. The closed EMC cabinet contained the γ-detector and all recording equipment. A 0.05 mm thick, 15 cm diameter aluminum window allowed the hard radiation quanta to pass. It also maintained sufficient shielding against the surge electromagnetic interference. This was demonstrated first by a small number of HV surges, when no hard radiation was detected and only the noise level of the oscilloscope was recorded. Second, we used three different scintillators with response times between 230 and 4 ns, and all detectors produced signals in the HV surge measurements with waveforms similar to those obtained for individual gamma quanta from e.g. a [137]Cs γ-source, outside the HV lab. Our change of scintillators is equivalent to the comparison of detector output with and without scintillator as applied in [6,13]. Third, insufficient cabinet shielding or power supply surges usually lead to an oscillating waveform. This has never been observed with the oscilloscope sensitivity used in our measurements.

The 8.5 m distance between the Marx generator and the spark gap reduced the chance that the detector captured hard radiation from the twelve generator spark switches.

A Tektronix TDS 3054 four-channel 8 bit digital oscilloscope recorded the HV divider output after further reduction by a factor of 40. The scope also registered the current through the grounded electrode via a Pearson 110 current probe with a rise time of 20 ns. The probe was mounted near the floor, at the grounded end of a 2 m long wire to the electrode. The γ-detector output was most often fed into two channels with a factor of 10 different in sensitivities in order to enhance the dynamic range. After a HV surge all data were automatically saved on a computer. This allowed uninterrupted measurements, and many hundreds of surges have been recorded.

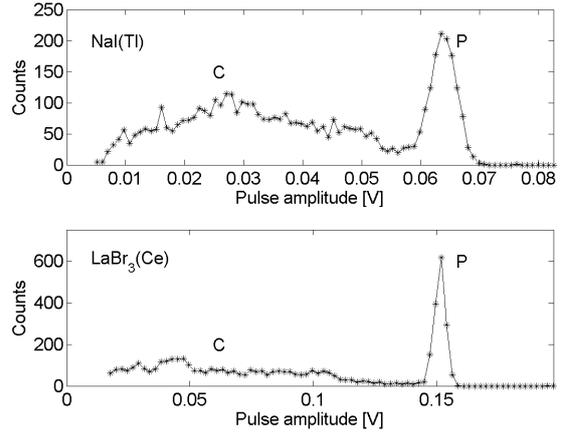

**Figure 2:** [137]Cs pulse height spectrum for NaI(Tl) and LaBr3(Ce) detectors of 5000 pulses recorded by an 8 bit resolution oscilloscope. The horizontal axes have been scaled to coincide on the photopeak.

*2.1 Gamma detectors*

We used three types of γ-detector with NaI(Tl), LaBr$_3$(Ce), and BaF$_2$ scintillator crystals. All scintillators were attached to photomultipliers with adequate speed. In Figure 2 we compare the first two materials. The γ-source was a sample of [137]Cs emitting characteristic γs of 662 keV. A total number of 5000 pulses have been recorded without further signal processing by an oscilloscope with 8 bit amplitude resolution. The pulse peak values were determined, and we plotted in Figure 2 the number of occurrences in bins of 1 resp. 2 mV versus the peak values. The spectra show that a direct interpretation of photomultiplier output pulse height in terms of incoming γ-quantum energy is not allowed. A γ-quantum can be absorbed completely in the scintillator and is then detected in the photopeak P. However, it is more likely that the γ-quantum undergoes Compton scattering. When the scattered γ-quantum escapes the scintillator the output signal will be correspondingly smaller as shown by the large and broad Compton ridge C.

**Table I**
Characteristics of the three scintillator materials in a Philips PW4119 detector, a Brilliance 380 detector by St. Gobain [17] and a BaF$_2$ detector by Scionix.

|   | NaI(Tl) | LaBr$_3$(Ce) | BaF$_2$ |
|---|---|---|---|
| a) # Photons/keV | 38 | 63 | 1.8 |
| b) Rise/fall [ns] | 40/230 | 11/23 | – |
| c) FWHM [ns] | 270 | 38 | 4 |
| d) En. resolution | 7.8% | 3.3% | 12% |
| e) Compton/Photo | 7 | 4 | – |
| Provider | Philips | St. Gobain | Scionix |

Several characteristics of three materials are summarized in Table I: a) the number of optical photons per keV absorbed γ-energy, b) the rise and fall times of the output pulse, c) the full width in time at half height of the light output. The width at half height of the photopeak in an amplitude spectrum for $^{137}$Cs γs is given in row d), now recorded [17] with adequate waveshaping electronics and a multichannel analyser. Row e) shows the probability ratio of detection in the Compton ridge or in the photo peak.
Introductory measurements were performed with the NaI(Tl) detector. It was quickly superseded by the modern LaBr$_3$(Ce) detector because of its faster response and better energy resolution. The 0.5 mm thick aluminum front of the scintillator encasement reduces the detection efficiency below 17 keV. For the measurements discussed here the time resolution of the LaBr$_3$(Ce) detector is about 4 ns on the 11 ns leading edge. A few measurements have been taken with a BaF$_2$ detector. The signal of this detector is composed of a fast component of 4 ns duration and a slow component. We only regarded the fast component.
Because most of the data to be presented have been obtained with the LaBr$_3$(Ce) detector, we determined its response in more detail. A model pulse waveform was obtained from $^{137}$Cs 662 keV radiation. In order to reduce digitizing noise, we averaged the records of 1200 pulses with amplitudes inside a window of 10% around the photopeak. This averaged waveform was then available as a numerical time series. It included the response of the photomultiplier, and deviated substantially from the bi-exponential waveform which suited well for NaI(Tl); see also Section 3.1. With this model waveform we determined the equivalent energy of the X-ray pulses from a HV surge, by adjusting time and amplitude in a least square fit procedure. This implicitly assumes that the response does not depend on energy or signal amplitude. The fit was even successful when the signal was slightly clipped by the oscilloscope, leaving the clipped data out of the fit. The procedure has been verified by using data recorded simultaneously on two channels with different sensitivity, with one dataset clipped. For appreciably broadened X-ray pulses we fitted the data to a train of model pulses with the smallest number of pulses possible. An example of such a train fit is presented in Section 3.3. Hereafter all amplitudes are expressed in equivalent radiation energy using the $^{137}$Cs calibration. But again, we caution for a direct interpretation of signal amplitude into equivalent energy neglecting Compton scattering. On the other hand, it may also occur that several quanta are absorbed within the response time of scintillator and photomultiplier. Pile-up of their signals then occurs.
A remark about wording: gamma rays are commonly associated with nuclear processes, X-rays involve electrons. Hard X-rays and soft gammas overlap in energy. We retained the term γ-detector because of the usually intended application.

---

[2] ElectroMagnetic Compatibility.

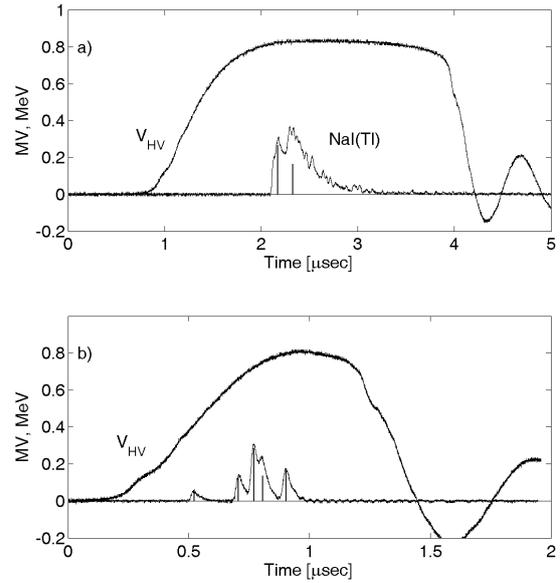

**Figure 3:** a) Positive HV surge with 830 kV maximum and the signal from the NaI(Tl) detector shown inverted, with bars indicating the fit amplitude and time. b) An 830 kV positive surge (V$_{HV}$), with the signal from the LaBr$_3$(Ce) detector inverted. Five pulses are distinguished, with equivalent energy up to 0.28 MeV. Please note the difference in time scale with respect to panel a). The longer pulse duration in panel a) stems from the larger gap distance: 1.2 m for a) and 0.7 m for b).

## 3. Experimental data
### 3.1 Comparison of γ-detectors
We first present two measurements to compare the detectors most used. The NaI(Tl) detector recorded signals in 50 percent of the positive HV surges, but much less for negative polarity. Figure 3a shows an example of a measurement: the surge voltage V$_{HV}$ measured by the HV divider together with the simultaneous record by the NaI(Tl) detector. The X-ray signal is displayed inverted for convenience. The gap electrode distance was 1.20 m. This record is similar to those published in [13]. Our HV surge started at time t = 0.75 μs, reached the maximum at t = 2.50 μs and collapsed due to the spark gap breakdown at t = 3.85 μs, and then developed in a damped oscillation. The response of the γ-detector showed two barely resolved peaks at t = 2.18 and 2.30 μs. The response could be fitted to within the noise by two bi-exponential model pulses with rise and fall time constants shown in Table I. The vertical bars in Figure 3 indicate the value and time of the maximum of the model pulses. Fitted equivalent amplitudes corresponded to energies of 0.27 and 0.17 MeV respectively. One observes that the second pulse starts in the decay time of the first.
Figure 3b shows an early measurement with the LaBr$_3$(Ce) detector. The electrode distance was 0.7 m, which led to a shorter time to breakdown than in Figure 3a. The solid angle Ω of the detector is about

$1.4\times10^{-3}$ sterad as viewed from the electrodes or developing spark. Five X-ray pulses can be recognized with equivalent energies of 0.05, 0.13, 0.28, 0.14 and 0.16 MeV in order of occurrence.

For the measurement with the NaI(Tl) detector, the electrodes were covered by a thin lead foil. For the LaBr$_3$(Ce) detector, the electrodes were aluminum. No difference between the radiation production was observed, as will be discussed in Section 4.

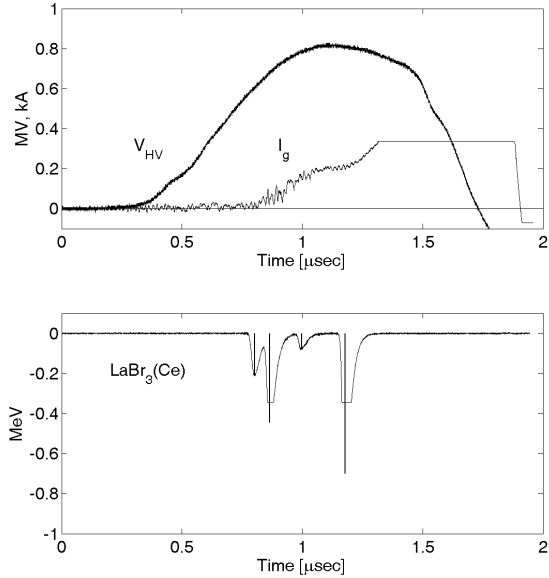

**Figure 4:** Positive surge with 810 kV maximum, shown with the current $I_g$ during the leader initiation phase. At 1.5 µs the gap breaks down. Four X-ray pulses can be distinguished. The γ-detector and current signals are clipped by the oscilloscope. The vertical bars indicate the maximum equivalent energy and its timing obtained by the fit.

*3.2 Positive HV surges*
We made two runs of 25 HV surges with 0.8 to 1.0 MV positive on the floating electrode. In all 50 surges, hard radiation has been observed with the LaBr$_3$(Ce) detector. Four X-ray pulses with equivalent energies up to about 700 keV can be recognized in the example shown in Figure 4. The amplitude and time of the X-ray pulses again resulted from the fit to the model pulse. The top part includes the current $I_g$ through the grounded electrode. The leader current started at t = 0.8 µs and the gap broke down completely at t = 1.5 µs. This time interval coincided with the detection of the X-rays. With this current range, the capacitive current that charges the spark gap electrodes (before t = 0.8 µs or before leader formation) is too small to be resolved.

The majority of X-rays occurred after about 75% of $V_{max}$. In contrast to [13] none occurred at the start of the surge or at gap breakdown. The timing and the current behavior links the X-ray production to the leader formation. For most positive surges the equivalent energy per X-ray pulse was smaller than $eV_{max}$. These pulses could be fitted to the model pulse down to the noise level. This indicates that these pulses corresponded either to a single X-ray quantum, or to the simultaneous detection of several lesser energy quanta well within the 4 ns detector time resolution. One of the surges produced an X-ray pulse with equivalent energy of 3 MeV; it appeared slightly broadened in time. Since it is hard to imagine that a single 3 MeV quantum is produced in our 1 MV discharges, we favor the interpretation in terms of a pile-up of several quanta within the detector time resolution. In line with this interpretation, we will use the term 'burst' rather than 'pulse' hereafter.

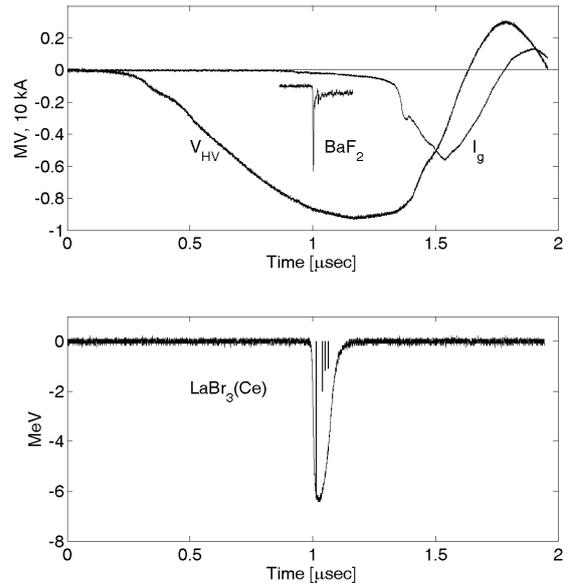

**Figure 5:** A negative 875 kV surge, shown with the current $I_g$ and X-ray signal (arbitrary units) from an uncalibrated BaF$_2$ detector alongside the LaBr$_3$(Ce) detector. The lower part of the figure shows the signal of the LaBr$_3$(Ce) detector and the fitted amplitudes of a train of four pulses.

*3.3 Negative HV surges*
With the same setup negative HV surges always produced much stronger X-ray signals, with peak equivalent energy per burst of several MeV, a few even up to 30 MeV as determined from the maximum detector output. Figure 5 shows an example with peak value equivalent to 6 MeV. The distance between cabinet and arc was 0.9 m. With 76 ns FWHM the LaBr$_3$(Ce) signal is significantly broadened compared to the model waveform (38 ns, see Table I). Also the peak is flattened appreciably. In a series of 20 surges 14 showed a LaBr$_3$(Ce) signal with averaged peak value of 5.8±1.3 MeV, all significantly broadened in time. The broadening may be attributed to a) afterglow in the scintillator, b) saturation of the photomultiplier or c) distribution in time of the X-rays. Cause a) is unlikely because of the decay time data presented in [18]; so there remains a combination of b) and c). We reduced the voltage of the photomultiplier and thereby its gain. The broadening remained. We fitted a time series of model pulses to the measured X-ray signal. The result is included in Figure 5 by the vertical bars indicating values and times of the maxima of the individual model pulses. In an attempt to resolve these strong signals further in time, we had installed a fast $BaF_2$ detector directly next to the $LaBr_3(Ce)$. This detector was only available for a short period and was not calibrated. The top part of Figure 5 also shows the record for the $BaF_2$ detector as inset on the same time scale. The single fast response of $BaF_2$ on the burst coincided with the onset of the $LaBr_3(Ce)$ signal. This was also observed in 12 out of the 14 surges mentioned before. Consequently we favor the interpretation that the large $LaBr_3(Ce)$ signal is stretched in time due to saturation of the photomultiplier. The sum of the amplitudes obtained from the fit is a lower limit for the scintillation light seen by the photomultiplier.

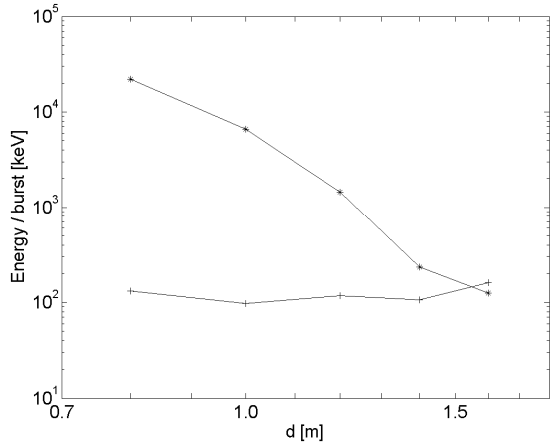

**Figure 6:** Energy per burst for positive (+) and negative (*) discharges, as a function of distance $d$ between cabinet and spark gap.

*3.4 Distance variation*
The large differences between X-ray production of the positive and negative surges was not compatible with the symmetry of the spark gap. In order to gain insight in where the X-rays were produced, the gap was placed at different distances from the EMC cabinet ($d$ in Figure 1) and a few tens of surges were produced at each position. The fit procedure gave the total equivalent energy per burst as sum of the fitted amplitudes. Surges where multiple bursts could be recognized were also analyzed this way. Figure 6 shows the result. In case of negative surges the energy per burst decreased rapidly for larger distances, approximately proportional to $d^{-7}$, and approached the value for positive surges at $d = 1.5$ m. For positive surges the variation was much less, if any at all. For a point-like source a $d^{-2}$ behavior would be expected. A line-like source would rather show to be proportional to $d^{-1}$.

Careful inspection by unaided eye revealed that for small $d$ arc initiation took place on the EMC cabinet with negative surge polarity, often on a ring holding the aluminum window. Most likely, X-rays were produced also there, right in front of the detector. No such phenomena were observed for larger distances $d$ or positive surges. This is in agreement with the X-ray abundance in negative surges at small $d$. Full breakdown to the cabinet occurred seldom at any $d$.

*3.5 Absorber*
A few measurements on negative surges have been taken with two $LaBr_3(Ce)$ detectors, placed alongside in the EMC cabinet. One detector was fully wrapped in a 1.5 mm thick lead foil. This foil provides a 1/e cutoff at 1.4 MeV, derived by the mass absorption coefficient from the NIST database [19]. It should be noted that these coefficients include all scattering mechanisms and assume single energy quanta and a monochromatic detector. Our detectors are not tuned for a single energy and will also record lower energy quanta emerging from the absorber after Compton scattering.

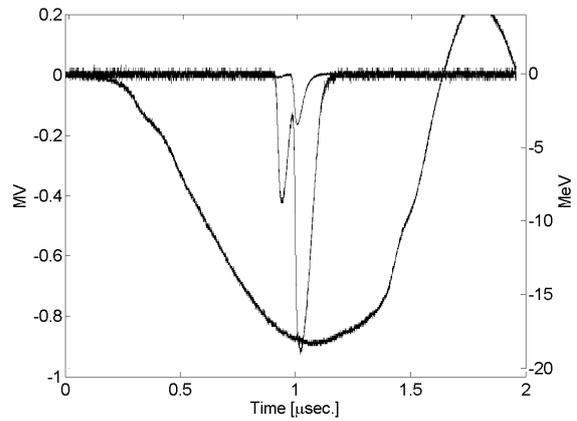

**Figure 7:** Negative surge (left ordinate) and X-ray signals (right ordinate) seen by two detectors, one without (large signal) and one with (smaller signal) a 1.5 mm Pb-absorber.

Figure 7 shows the results for a 0.88 MV surge. The distance $d$ was 0.9 m. The peak value of the largest X-ray burst at t = 1 μs was equivalent to 19 MeV for the γ-detector without absorber; again the signal was widened in time to a FWHM of 79 ns. The total equivalent energy obtained from a fit was 43 MeV. For the one wrapped in lead the largest burst corresponded to 3.5 MeV; it was not appreciably widened. The total energy was 4.7 MeV. The strong reduction of the signal by the lead agrees with the fact that the larger signal consists of a pile-up of many lesser energy X-ray quanta. If one assumes that quanta are evenly distributed in space and energy over both detectors, and one assumes that only the high energies contribute to the burst signals, the observed ratio of the signals 43/4.7 = exp(2.2) can be converted into a mass absorption coefficient μ of 1.3 cm$^2$/g. With the NIST XCOM tables [19], the value of μ corresponds to quantum energies between 90 and 150 keV. For the smaller burst at t = 0,9 μs a similar analysis leads to μ = 2.2 cm$^2$/g and an energy of about 80 keV. The energies are not very sensitive to variations of µ.

## 4. Discussion and Conclusion

X-rays bursts have been observed for positive and negative surges. We limited the field of view of a $LaBr_3(Ce)$ detector by a lead tube to separate ranges around the HV electrode, around the grounded electrode and on the space midway in between. In case of positive surges we only detected X-rays looking in the direction of the HV electrode. In case of negative surges, the EMC cabinet acted as an additional positive electrode and contributed strongly to the X-ray production if the distance was small enough, see Figure 6. As a result, measurements with field of view limitation were not conclusive. Still, for both surge polarities the majority of observed X-rays originated near the positive electrode.

If the X-rays are formed by collisions of high-energy electrons at the anode, one would expect an increase if the aluminum surface was covered by lead. This has been tried, but no substantial increase was found.

The bursts coincide in time with the large current rise on the grounded electrode, indicative for leader formation. No X-ray signals have been observed at or after the sharp rise of the current at full breakdown. The multiple bursts indicate that the discharge develops through a stepped streamer/leader formation process. However, the signal to noise ratio, the time-resolution and position of the present current probe did not allow to distinguish corresponding steps in the currents.

We measured the current through the grounded electrode, which is the negative side of the spark gap for positive surges. It is remarkable that the X-rays then occur only when the current starts to rise to few 100 A, whilst the primary leader forms at the (positive) HV electrode. Future experiments will include current measurements on both electrodes.

The X-ray signal is not directly related to the energy of the quanta arriving at the scintillator. It is hard to imagine quanta with energy larger than $eV_{max}$, except if a streamer ionization wave moves with the same velocity as runaway electrons in the streamer head [10,20]. The observed multi-tens of MeV signals are a pile-up of many lesser energy quanta. The few measurements with the lead absorber indicate that the bursts contain hard quanta of the order of 100 keV. Further absorber measurements are under way.

The widening of the very intense bursts has been analyzed as a train of model pulses to determine the total signal intensity. The amplitude of the first pulse was always largest, see for instance Figure 5. Comparison of the signals with those from a $BaF_2$ detector points at saturation effects in the $LaBr_3(Ce)$ photomultiplier.

For negative surges the total energy of hard radiation strongly depended on the distance between the spark gap and the EMC cabinet, and discharge initiation was seen at the cabinet if close enough to the discharge gap. As a result, the possibility should be recognized that X-rays originate not only from the spark gap, but can also be formed elsewhere if the local electric field is large enough for discharge initiation. This possibility should also be considered for outdoor measurements. For positive surges, no such secondary discharges were seen which agrees with the observation that negative streamers are much harder to initiate from metal electrodes than positive ones [21]. The electric field conditions are most likely met at the heads of streamers emanating from the HV electrode.

The average total X-ray energy from positive surges is of the order of a few hundred keV. A single quantum may be responsible for the signal, or at most a few tens of quanta taking the 17 keV lower detection limit into account. Assuming isotropic emission and taking the detector solid angle $\Omega = 1.4 \times 10^{-3}$ sterad into account, one deduces that at least $4\pi/\Omega \approx 10^4$ quanta and electrons with a few hundred keV contribute to each X-ray burst, and several times this number for lesser energy quanta. This is a large number in view of current theoretical models for the electron energy distribution [10, 20, 22-24] in the developing discharge, in particular if one takes into account that the electrons causing this emission are in the extreme high energy tail of the electron energy distribution. We will further investigate the angular distribution with two $LaBr_3(Ce)$ detectors.

The recent paper [14] also discusses X-ray production during similar discharges of a lightning surge generator, voltages of the order of 1 MV, spark gap distance of about 1 m. Their $BaF_2$ detector was mounted at about 1 m from the spark gap in a floating shielded cabinet. In contrast to our results only X-rays have been detected during negative surges, and also X-rays have been seen at the moment of full gap breakdown. Even with similar slightly asymmetric gaps (grounded electrode rounded disk, 8 cm diameter) we observed X-rays for both polarity. The multi-bursts shown for instance in Figure 3 were not reported. Both experiments show that experimental research on run-away electrons in long sparks can be performed in the laboratory. The signals are a multiple integral over the X-ray's energy distribution, folded with the detector response, and their direction in space and timing folded with the detector characteristic times. To unravel the signals into an electron energy distribution function requires substantial effort.

## Acknowledgement
The authors thank P. van Rijsingen of the Radboud University in Nijmegen (NL) for the loan of the NaI detector.